\documentclass[%
 aip,
 jmp %
 amsmath,amssymb,
 reprint,%
]{revtex4-1}

\usepackage{graphicx}
\usepackage{dcolumn}
\usepackage{bm}

\begin{document}

\preprint{AIP/123-QED}

\title{Determination of the Optical Index for Few-Layer Graphene by Reflectivity Spectroscopy}

\author{Behnood G. Ghamsari}
\email[Electronic mail: ]{ghamsari@umd.edu}
\author{Jacob Tosado}
\author{Mahito Yamamoto}
\author{Michael S. Fuhrer}
\author{Steven M. Anlage}
\affiliation{\footnotesize Center for Nanophysics and Advanced
Materials, Department of Physics, University of Maryland, College
Park, MD, 20742-4111, USA}

\date{\today}

\begin{abstract}
We have experimentally studied the optical refractive index of few-layer graphene through reflection spectroscopy at visible wavelengths.
A laser scanning microscope (LSM) with a coherent supercontinuum laser source measured the reflectivity of an exfoliated graphene flake on a Si/SiO$_2$ substrate, containing monolayer, bilayer and trilayer areas, as
the wavelength of the laser was varied from 545nm to 710nm. The complex refractive index of few-layer graphene, $n-ik$, was extracted from the reflectivity contrast to the bare substrate and the Fresnel reflection theory. Since the SiO$_2$ thickness enters to the modeling as a parameter, it was precisely measured at the location of the sample.
It was found that a common constant optical index cannot explain the wavelength-dependent reflectivity data for single-, double- and three-layer graphene simultaneously, but rather each individual few-layer graphene possesses a unique optical index whose complex values were precisely and accurately determined from the experimental data.
\end{abstract}

\pacs{Valid PACS appear here}


\maketitle

\section{Introduction}

The recent isolation of graphene has inspired tremendous efforts to investigate the physical properties of few-layer graphene, as well as their applications\cite{Geim,Bonaccorso,Schwierz}.
The visibility of few-layer graphene on common dielectric substrates under ambient conditions has been used as an effective and convenient means for locating and identifying different graphene flakes through their reflectivity contrast to the substrate \cite{Blake, Ni}, provided that the substrate and the wavelength of the light are properly chosen. More sophisticated and definitive methods, such as Raman spectroscopy and atomic force microscopy (AFM), have been also frequently used to quantify the number of graphene layers\cite{Ferrari}.

Aside from its immediate role as an experimental technique, the visibility of sub-nanometer-thick graphene flakes implies strong interaction between the graphene electronic system and visible/near-infrared light, and highlights the potential of graphene as an optoelectronic material.
Insofar as the refractive index of any optoelectronic material is one of its most fundamental optical properties, many groups\cite{Blake,Ni,Nair,Mak,Wang,Skulason,Bruna,Gaskell,Teo,Lenski} have attempted to probe the effective optical index of few-layer graphene. In these works, the Fresnel theory of reflection/refraction has been widely applied to interpret experimental data as well as devising predictions on observable quantities like absorption, transmission, and reflectivity contrast. Clearly, these quantities are important to a broad range of photonic applications such as graphene photodetectors, solar cells, transparent electronics, and phototransistors.

Nevertheless, the quantitative value of the effective refractive index for graphene, which enters to the Fresnel theory as a constitutive parameter, greatly varies among various reports. The proposed values range from the optical index of bulk graphite\cite{Blake}, $n_G=2.6-1.3i$, to an optimized value based on the best fit to reflectivity data\cite{Ni}, $n_G=2.0-1.1i$; whereas transmission measurements support a universal optical conductivity picture\cite{Nair,Mak}, $G\equiv\sigma\cdot d=e^2/4\hbar$, with $\sigma$ being the three dimensional conductivity, $d$ the graphene thickness, $e$ the electron charge and $\hbar$ Planck's constant, and predicts 2.3\% of absorption per layer of suspended graphene. There are also some reports and conjectures on the possibility of optical dispersion in few-layer graphene at visible and near-infrared wavelengths\cite{Wang,Skulason}. While much of the related experimental work has been on exfoliated samples, similar experiments on graphene samples prepared by chemical vapor deposition (CVD) have also yielded qualitatively similar results\cite{Lenski}.

The controversy about the value of the complex optical index for graphene, its possible variation with the number of layers and dispersion with the wavelength, and the hypothesis of a constant optical conductivity well up to the visible regime have inspired this work to experimentally revisit the frequency-dependent optical behavior of few-layer graphene, including monolayer, bilayer, and trilayer, in the visible regime. Here we have applied an improved method of reflectivity spectroscopy using coherent light to enhance the sensitivity of the experiment to both the imaginary and real parts of the optical index, especially since the latter effectively plays no role in transmission experiments.

For precise and accurate determination of the optical index, reflectivity measurements have a significant advantage over transmission experiments.
Given the optical length of the whole structure is vanishingly small compared to the wavelength, the transmittance/opacity of suspended graphene samples is predominantly determined by the absorption of light waves in the graphene sample, since optical beams pass through the flake effectively once and the one-time reflection at the air-graphene boundary is negligible and in many instances beyond the sensitivity of the measuring apparatus.

In contrast, experimental determination of the complex optical constant, $n-ik$, through reflectivity of a graphene flake on a dielectric substrate provides higher sensitivity to the real part of the optical index, $n$, and is capable of probing the complex optical index more accurately.  If the graphene flake is coupled to an optical cavity, such as a Si-SiO$_2$ Fabry-Perot resonator, the effect of $n$ is accentuated because the reflection will be an Airy superposition over all the resonating rays inside the cavity that are interfering with each other according to their various phases and optical paths, thereby, carrying a strong contribution from $n$.

While the great majority of earlier reflectivity experiments have used incoherent light and a combination of high numerical-aperture (NA) objective lenses or image processing techniques to extract the refractive index of graphene, a confocal laser scanning microscope (LSM) with a coherent light source together with a moderate NA objective lens have been utilized in this work to improve the accuracy and precision of our measurements.

\section{Experiment}

A graphene flake was exfoliated onto a commercially available Si substrate with a nominally 300nm-thick SiO$_2$ over-layer.
The sample is then annealed in H$_2$ and Ar at a temperature of 500$^\circ$C for 3 hours.
Regions of mono-layer, bi-layer, and trilayer were identified by means of Raman spectroscopy and atomic force microscopy (AFM), the latter was performed after the reflectivity measurements
to prevent possible damage to the surface.

\begin{figure}
\includegraphics[width=3.5in]{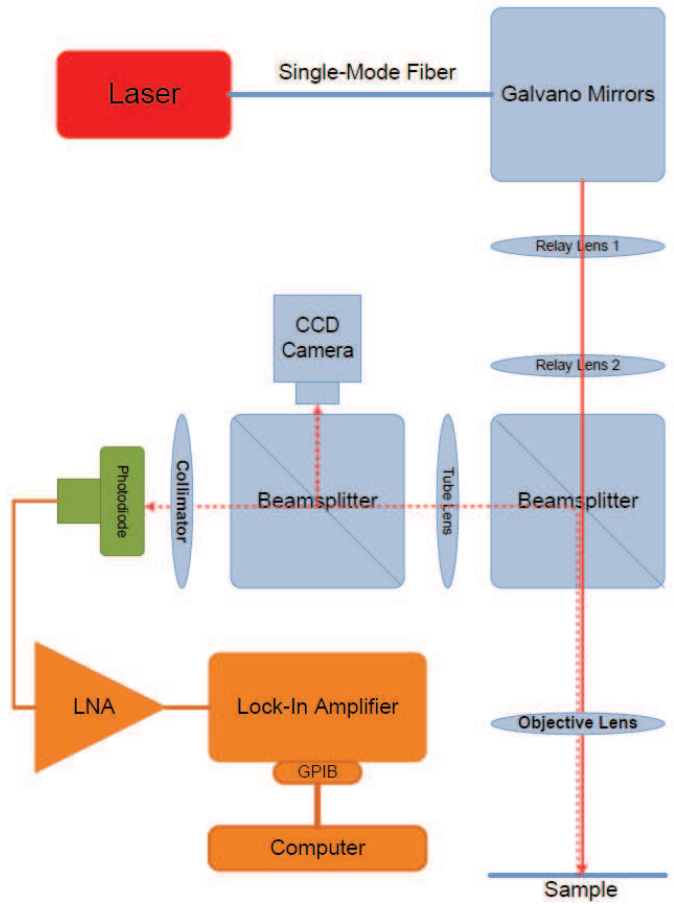}
\caption{\label{Setup}Schematic of the Laser Scanning Microscope.}
\end{figure}

\begin{figure}
\includegraphics[width=3.5in]{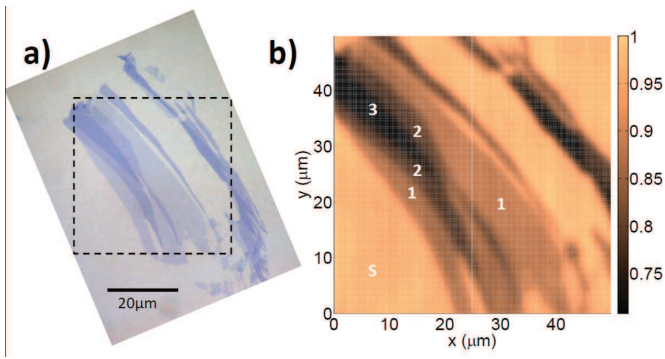}
\caption{\label{Fig1}a) Optical image of the sample containing monolayer, bilayer, and trilayer graphene. The dashed-line square signifies the LSM field of view. b) LSM reflectivity image of the sample at 638nm.
The amplitudes are normalized to the maximum in the image. The S, 1, 2, and 3 signs respectively label regions of the substrate, monolayer, bilayer, and trilayer.}
\end{figure}

Reflectivity measurements were performed by a laser scanning microscope (LSM)\cite{Zhuravel}, as illustrated in Figure \ref{Setup}, which raster-scans a focused laser spot of nominally 1$\mu m$ diameter over the sample using a set of two galvano-mirrors. An X20 Mitutoyo infinity-corrected long working-distance objective lens, with a NA of 0.42, was used to focus the light onto and to gather the reflected light from the sample.
The reflected light was directed to a Si photodiode which measured the reflected light at each pixel.
Figure \ref{Fig1} shows the optical image of the graphene flake as well as the corresponding LSM reflectivity image at a wavelength of 638nm.

\begin{figure}
\includegraphics[width=3.5in]{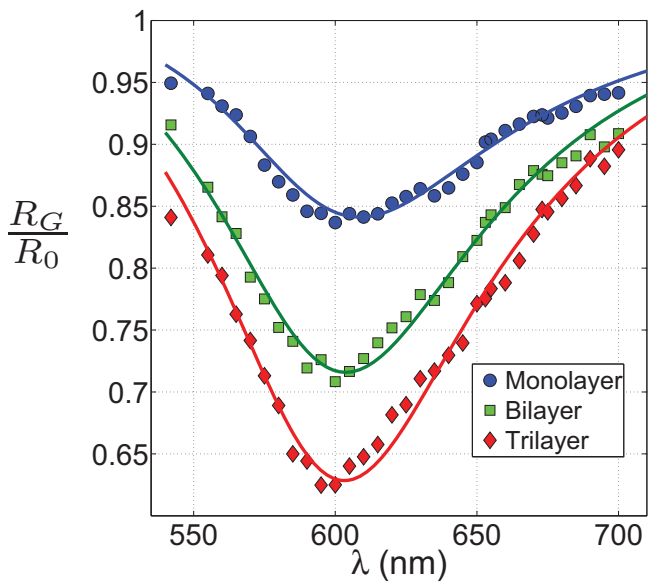}
\caption{\label{Fig2}The normalized reflectance of different few-layer graphene regions as a function of wavelength.
Normalized reflectance is defined as the ratio of the graphene's reflectance to that of the bare substrate at the same wavelength.
Circles, squares, and diamonds respectively show the experimental data for monolayer, bilayer, and trilayer graphene and the solid lines are the best fit based on the Fresnel theory for each individual graphene layer irrespective of the others.}
\end{figure}

A Fianium SC400 supercontinuum laser was used as the LSM light source whose wavelength was varied from 545nm to 700nm by means of an acousto-optic tunable filter with a bandwidth of $<$2nm.
At each wavelength, the intensity histogram of the LSM reflectivity image showed four peaks corresponding to the bare substrate, monolayer, bilayer, and trilayer regions.
By fitting a Gaussian distribution to each peak and taking the mean value to represent the reflectivity of the corresponding region, the frequency-dependent reflectivity for different graphene thicknesses were obtained, as Figure \ref{Fig2} demonstrates.

The thickness of the oxide layer enters to the data analysis as an input parameter and its accurate determination is important to the accuracy of the ultimate results.
Therefore, although ellipsometry and scanning electron microscopy (SEM) were performed on the substrate, the actual thickness near the graphene flake was determined, after the reflectivity measurements, through
selectively removing the oxide by hydrofluoric acid etching followed by AFM, which gave a value of 308nm$\pm$0.5nm for the oxide thickness adjacent to the exfoliated graphene flake.

\section{Analysis and Results}

The Fresnel theory\cite{Yeh} of reflection is often applied in order to interpret reflectivity/transmission experiments on graphene.
Here, we have also used the same method to analyze our data including the effects of dispersion of Si and SiO$_2$ in the optical band of interest\cite{Palik}.

A dispersionless optical index is the simplest assumption for the optical behavior of graphene, and has been widely used among many researchers in the field\cite{Blake,Ni,Teo,Lenski}.
Although the Kramers-Kronig theorem states that a dissipative material must be dispersive as well, these two do not have to necessarily occur over the same band simultaneously, and it is perfectly admissible to presume a dispersion free index over a limited frequency band.
It should be noted, however, that this assumption is not equivalent to a universal optical conductivity\cite{Kuzmenko,Bruna}, $\sigma$, which is a function of both $n$ and $k$, since $\epsilon\equiv(n-ik)^2=\epsilon^\prime - i\sigma/\epsilon_0\omega$, where $\epsilon$ is the relative dielectric constant, $\epsilon_0$ is the permittivity of vacuum, $\epsilon^\prime \equiv\Re e\{\epsilon\}$, and $\omega$ is the angular frequency of light.

It has been also conjectured that the optical response of few-layer graphene is predominantly determined by the in-plane electrodynamics\cite{Blake} and, thus, the optical indices of few-layer graphene should be similar.
We also put this assumption to the test and try to extract refractive indices for few-layer graphene independently and then examine the aforementioned hypothesis.

With these assumptions, the optical index of graphene enters into our model as a constant complex number whose value is found based on the best fit to the $R_G/R_0$ vs. wavelength experimental data for each few-layer graphene individually. The obtained complex optical indices through this algorithm are tabulated in Table \ref{nk} and the theoretical reflectivity curves associated with them are depicted in Figure \ref{Fig2}.

Clearly, a dispersionless optical index very well models the experimental reflectivities, nonetheless, the optical indices for monolayer, bilayer, and trilayer graphene are different.
Given that all the few-layer graphene areas in this experiment have been produced and measured simultaneously and, thereby, under essentially identical conditions, the discrepancy among the optical indices is not a consequence of measurement uncertainties, but rather suggests that each few-layer graphene possesses a unique optical index.

We also examined the dispersive model of Bruna and Borini \cite{Bruna} to model the experimental data, however it did not lead to a satisfactory fit for any $n$ between 2.0 and 3.2, except for the monolayer sample in the low energy range of $\lambda>600nm$.

\begin{table}
\caption{\label{nk}Dispersionless optical index of graphene obtained from the reflectivity spectroscopy.}
\begin{ruledtabular}
\begin{tabular}{lcr}
Graphene &  n & k\\
\hline
Monolayer & 2.69$\pm$0.02 & 1.52$\pm$0.02\\
Bilayer & 2.38$\pm$0.02 & 1.66$\pm$0.02\\
Trilayer & 2.27$\pm$0.02 & 1.60$\pm$0.02\\
\end{tabular}
\end{ruledtabular}
\end{table}

Nevertheless, as Table \ref{nk} shows, the imaginary parts are nearly equal; besides, the majority of transmission/absorption experiments on graphene are not as sensitive to the real part of the optical index, simply because they use  suspended graphene samples in which no standing-wave pattern forms due to its small thickness compared to the wavelength. These two reasons have enabled interpreting graphene transmission/absorption experiments based on a common constant optical conductivity in the visible regime.

\section{Conclusions}

We have performed reflectivity spectroscopy on exfoliated monolayer, bilayer, and trilayer graphene by means of a laser scanning microscope and a supercontinuum laser over the wavelengths of 545nm to 710nm, corresponding to photons of energy 1.75eV to 2.28eV.
The optical index of the few-layer graphene flakes were determined based on their contrast to the bare Si-SiO$_2$ substrate by applying the Fresnel theory of reflection/refraction to the air/graphene/SiO$_2$/Si multilayer.
We found that the experimental reflectivity data for each few-layer graphene  can be very well explained by a constant dispersionless effective optical refractive index rather than a constant optical conductivity; however, the indices are distinct for each of the few-layers and the optical conductivity does not scale with the number of layers. As listed in table \ref{nk}, the extinction factor, $k$, for monolayer, bilayer, and trilayer graphene are nearly equal, nonetheless, their real parts, $n$, are significantly different.

\begin{acknowledgments}
B.G.G. would like to thank Fianium for generously providing the supercontinuum laser system for the course of the reflectivity experiments as well as Sten Tornegard and Tim Gerke for technical support.
B.G.G. and J.T would like to thank Claudia Ojeda-Aristizabal for providing advice in the SiO$_2$ etching.
This work is supported by the DOE grant number DESC0004950 and ONR/UMD/AppEl, Task D10, through grant number N000140911190.
\end{acknowledgments}

\end{document}